\documentclass[11pt]{amsart}
\usepackage{latexsym,amsfonts,amsthm,amssymb, epsfig}
\newtheorem{thm}{Theorem}

\newtheorem{prop}[thm]{Proposition}
\newtheorem{cor}[thm]{Corollary}

\theoremstyle{remark}
\newtheorem{rem}{Remark}

\theoremstyle{definition}

\newcommand{\R}{\mathbb{R}}

\newcommand{\N}{\mathbb{N}}

\newcommand{\Z}{\mathbb{Z}}

\newcommand{\calP}{\mathcal{P}}

\newcommand{\LL}{\mathbb{L}}

\DeclareMathOperator{\disp}{disp}

\title[Dispersion of admissible lattices]{A note on the dispersion\\ of admissible lattices}
\author{Mario Ullrich}
\date{\today}

\begin{document}

\begin{abstract}
In this note we show that the volume of axis-parallel boxes in $\R^d$ which do not 
intersect an admissible lattice $\LL\subset\R^d$ is uniformly bounded. 
In particular, this implies that the dispersion of the dilated lattices $N^{-1/d}\LL$ 
restricted to the unit cube is of the (optimal) order $N^{-1}$ as $N$ goes to infinity.
This result was obtained independently by V.N.~Temlyakov (arXiv:1709.08158).
\end{abstract}

\maketitle

Let $\LL\subset\R^d$, $d\in\N$, be an \emph{admissible lattice}
(in the sense of Skriganov~\cite{Sk94}), 
i.e.,  $\LL=T(\Z^d)$ for some $T\in\R^{d\times d}$ such that
\begin{equation}\label{eq:adm}
{\rm Nm}(\LL) \,:=\, \inf_{z\in\LL\setminus\{0\}}\, \prod_{j=1}^d |z_j| \,>\, 0.
\end{equation}
Such lattices play a crucial role in the geometry of numbers, see e.g.~\cite{Ca97,GL87} 
and the references therein, 
especially in the theory of Diophantine approximation.
Moreover, they are among the most important explicit constructions of 
point sets, which satisfy various \emph{uniform distribution properties}, like the 
optimal order of the discrepancy~\cite{Fr80,Sk94}.
%Only recently, such lattices 
They attracted quite a lot of attention in numerical analysis 
as the corresponding point sets seem to serve as an optimal and universal choice as 
nodes for corresponding cubature rules, 
see e.g.~\cite{Fr76,KN16,NUU16,Sk94,Te93,Te03,MU15,MU17,UU16}.
See also \cite{No15,NW10,Te03} for surveys on the state of the art in numerical integration.

Apart from the concepts and geometric quantities that are important for the above, 
there is an increasing interest in the dispersion of a point set. 
For $d\in\N$ and a point set $\calP\subset[0,1]^d$, 
the \emph{dispersion of $\calP$} is defined by
\[
\disp(\calP) \;:=\; \sup_{B\colon B\cap\calP=\varnothing}\, |B|, 
\]
where the supremum is over all axis-parallel boxes $B=I_1\times\dots\times I_d$ 
with intervals $I_\ell\subset[0,1]$, 
and $|B|$ denotes the volume of $B$.
This quantity was proven to be essential for various numerical problems 
including optimization~\cite{Niederreiter83}, approximation of high-dimensional 
rank-1 tensors~\cite{BDDG14,NR15} and, very recently, in the study 
of Marcinkiewicz-type discretization theorems~\cite{Te17a,Te17b,Te17c}. 
Moreover, algorithms for finding such a box with maximal volume, 
with a motivation coming from information theory, are also of some interest, 
see~\cite{DJ13,DJ13b,NLH84} and references therein.
In all these contexts, it is desirable to have tight bounds on the dispersion 
(clearly depending on $d$ and $\#\calP$), and to find explicit constructions 
of point sets satisfying these bounds.
The best bounds so far, with special emphasis on the dependence on the dimension $d$, 
can be found in the recent articles~\cite{AHR15,Kr17,Sos17,MU18,UV17}. 
In particular, the only presently known optimal 
(w.r.t.~the dependence on $\#\calP$) 
construction is based on $(t,m,d)$-nets, see~\cite[Section~4]{AHR15}.
Here, we will prove that admissible lattices also lead to optimal 
(w.r.t.~the dependence on $\#\calP$) constructions for the dispersion.
This was obtained independently by Temlyakov~\cite{Te17d}. 
However, we show that this is a quite direct corollary of the famous 
lattice point counting result of Skriganov~\cite{Sk94}, which seems 
to be of independent interest. 
%In any case, our proof is much shorter.

In this note the point sets under consideration are admissible lattices $\LL$ 
and we consider the (apparently larger) \emph{lattice-dispersion of $\LL$}, 
i.e., the quantity
\[
\disp^*(\LL) \;:=\; \sup_{B\colon B\cap\LL=\varnothing}\, |B|,
\]
where the supremum is over all axis-parallel boxes $B=I_1\times\dots\times I_d$ 
with intervals $I_\ell\subset\R$, 
and $|B|$ denotes again the volume of $B$.

First, we prove the following result.

\begin{prop}\label{prop:main}
Let $\LL\subset\R^d$ be a lattice that satisfies~\eqref{eq:adm}.
Then, 
\[
\disp^*(\LL) \,\le\, D_\LL \,<\, \infty,
\]
where $D_\LL$ depends on $\LL$ only by means of 
$\det(\LL)$ and ${\rm Nm}(\LL)$.
\end{prop}
\medskip

\begin{rem}
In the application of such results, particularly in numerical analysis, 
it is often necessary to switch between properties of the lattice 
$\LL=T(\Z^d)$ and the corresponding \emph{dual lattice} $\LL^*=(T^{-1})^\top(\Z^d)$.
Let us therefore note that $\LL$ 
%is admissible (see~\eqref{eq:adm}) 
satisfies~\eqref{eq:adm}
if and only if 
$\LL^*$ satisfies~\eqref{eq:adm}, see~\cite[Lemma~3.1]{Sk94}. 
\end{rem}

\bigskip

From this result we obtain a bound on the dispersion (in the unit cube) 
for dilated admissible lattices. 
For this, we introduce the notation 
\[
t^{-1} \LL:=\{(t_1^{-1}z_1,\dots,t_d^{-1}z_d)\colon z\in\LL\}
\] 
and
$[0,t]:=[0,t_{1}]\times\dots\times[0,t_{d}]$
for $t=(t_1,\dots,t_d)\in\R^d$ with 
%and define
$n(t):=\bigl|[0,t]\bigr|=\prod_{j=1}^d|t_j|>0$.

\begin{cor}\label{cor1}
Let $\LL\subset\R^d$ be a lattice that satisfies~\eqref{eq:adm}.
Then, 
\[
\disp\Bigl((t^{-1} \LL)\cap[0,1]^d\Bigr) \,\le\, \frac{D_\LL}{n(t)}.
\]
Moreover, for every $N\in\N$ there exist $t_N\in\R^d$ such that
$\calP_N:=(t_N^{-1} \LL)\cap[0,1]^d$ satisfies $\#\calP_N=N$ and
\[
\disp(\calP_N) \,\le\, \frac{C_\LL}{N}
\]
for some constant $C_\LL$ that
%and $\widetilde{C}(\LL)$ 
depends on $\LL$ only by means of 
$\det(\LL)$ and ${\rm Nm}(\LL)$.
\end{cor}
\medskip

\begin{proof}[Proof of Proposition~\ref{prop:main}]
We know from the work of Skriganov~\cite[Theorem~1.1]{Sk94} that, 
for $B_{x,t}:=x+[0,t]$, $x,t\in\R^d$, we have
\[\begin{split}
\Bigl|\#\bigl(\LL\cap B_{x,t}\bigr) - \frac{|B_{x,t}|}{\det(\LL)}\Bigr|
\,\le\, D'_\LL\, \ln\Bigl(2+n(t)\Bigr)^{d-1}
\end{split}\]
uniformly in $x$, where the constant $D'_\LL$ 
depends on $\LL$ only by means of $\det(\LL)$ and ${\rm Nm}(\LL)$.
If we now assume that the (shifted and dilated) box $B_{x,t}$ 
contains no point of $\LL$, we obtain that
\[
|B_{x,t}| \,\le\, \det(\LL)\,D'_\LL\, \ln\Bigl(2+n(t)\Bigr)^{d-1}.
\]
This easily implies that $|B_{x,t}| = n(t) \le D_\LL$ for 
some $D_\LL<\infty$ that only depends on $\det(\LL)$ and $D'_\LL$.

%Assume to the contrary, $\disp^*(\LL)=\infty$ would imply
%$\disp^*(t^{-1}\LL)=\infty$ for any $t=(t_1,\dots,t_d)\in\R^d$.
%Here, we use the notation 
%$t^{-1} \LL=\{(t_1^{-1}z_1,\dots,t_d^{-1}z_d)\colon z\in\LL\}$.
%In particular, this would imply the existence of 
%$x\in\R^d$ and a sequence of $t=(t_1,\dots,t_d)\in\R^d$ with 
%$n(t):=\prod_{j=1}^d|t_j|\to\infty$, 
%such that $\#\bigl((x+t^{-1} \LL) \cap[0,1]^d\bigr)=0$ and 
%therefore
%\[
%\Bigl|\frac{\#\bigl((x+t^{-1} \LL) \cap[0,1]^d\bigr)}{n(t)}-\vol_d\bigl([0,1]^d\bigr)\Bigr| 
%\,=\, 1.
%\]
%However, we know from the work of Skriganov~\cite{Sk94} that 
%\[
%\Bigl|\frac{\#\bigl((x+t^{-1} \LL) \cap[0,1]^d\bigr)}{n(t)}-\vol_d\bigl([0,1]^d\bigr)\Bigr| 
%\,\le\, C\, \frac{\log^{d-1}(n(t))}{n(t)},
%\]
%uniformly in $x$ and the chosen sequence.
%The right hand side goes to zero as $n(t)\to\infty$: a contradiction.
\end{proof}

\medskip

\begin{proof}[Proof of Corollary~\ref{cor1}]
We obtain from Proposition~\ref{prop:main} that under the given assumption 
$\disp^*(\LL)\le D_\LL<\infty$. Therefore, we obtain from the definitions 
and the homogeneity of the lattice-dispersion that
\[
\disp\Bigl((t^{-1} \LL)\cap[0,1]^d\Bigr)
\,\le\, \disp^*\Bigl(t^{-1} \LL\Bigr)
\,=\, \frac{\disp^*\bigl(\LL\bigr)}{n(t)}
\,\le\, \frac{D_\LL}{n(t)}.
\]

For the second part of the corollary, note that the existence of 
%an 
infinitly many 
$t_N$ with $\#\calP_N=N$ follows from the fact that every hyperplane 
$\{x\colon x_j=a\}$, $a\in\R$, 
contains at most one point of $\LL$, see e.g.~\cite[Remark~2.1]{Sk94}. 
Now note that also $N=\#\bigl(\LL\cap[0,t_N]\bigr)$, 
%where $[0,t_N]:=[0,t_{N,1}]\times\dots\times[0,t_{N,d}]$, 
and that \eqref{eq:adm} implies $n(t_N)>{\rm Nm}(\LL)$ for $N\ge2$.
We partition $[0,t_N]$ into at most 
\[
|[0,t_N]|/{\rm Nm}(\LL)+1=n(t_N)/{\rm Nm}(\LL)+1\le 2\, n(t_N)
\]
axis-parallel boxes with volume less than ${\rm Nm}(\LL)$. 
Each of these boxes contains at most one point from $\LL$.
Otherwise, two distinct points $x,y\in\LL$ in the box must satisfy 
$\prod_{j=1}^d|x_j-y_j|<{\rm Nm}(\LL)$: a contradiction to~\eqref{eq:adm}.
Therefore, we obtain $N\le 2\, n(t_N)$, which proves the result 
with $C_\LL=2 D_\LL$.\\
\end{proof}

\goodbreak

\end{document}